\documentstyle[sprocl]{article}

\def\be{\begin{equation}}
\def\ee{\end{equation}}
\def\bea{\begin{eqnarray}}
\def\eea{\end{eqnarray}}

\begin{document}
Solicited talk in {\it Data Analysis in Astronomy}, Erice 1996
\bigskip
\bigskip

\title{ARTIFICIAL NEURAL NETWORKS AS A TOOL
FOR GALAXY  CLASSIFICATION}

\author{OFER LAHAV}

\address{Institute of Astronomy, Madingley Road,
Cambridge CB3 0HA, UK}

\maketitle\abstracts { We describe an Artificial Neural Network (ANN)
approach to classification of galaxy images and spectra.  ANNs can
replicate the classification of galaxy images by a human expert
to the same degree of agreement as that between two human experts, to
within 2 $T$-type units.  Similar methods are applied to 
classification of galaxy spectra.  In particular,  Principal Component
Analysis of galaxy spectra can be used to compress the data, to
suppress noise and to provide input to the ANNs.
These and other classification 
methods  will soon be applied to the Anglo-Australian 
2-degree-Field (2dF) redshift survey of 250,000 galaxies.}

\section{Introduction}

The morphological classification of bright galaxies is still
mainly done visually by  dedicated individuals,
in the spirit of Hubble's (1936) original scheme
and its  modifications
(e.g. Morgan 1958, de Vaucouleurs 1959, 1991,  Sandage 1961, 
van den Bergh 1976).
It is remarkable that  these somewhat subjective
classification labels  for galaxies
correlate well with physical properties
such as colour and  dynamical properties.
However, one would like eventually to devise 
schemes of classification which can be related to the physical
processes of galaxy formation.
While  there have  been in recent years
significant advances in
observational techniques (e.g. telescopes, detectors and
reduction algorithms) as well as in  theoretical modelling
(e.g. N-body and hydrodynamics simulations),
galaxy classification remains a subjective area.
Galaxy classification is important for both practical
reasons of producing large catalogues for statistical and observational
programs, as well as for establishing
some underlying physics (in analogy with the H-R diagram for stars).
Moreover, understanding the morphology of galaxies at low redshift
is crucial for any meaningful comparison with
galaxy images obtained with the Hubble Space Telescope
at higher redshift (e.g. the Hubble Deep Field).

Most of our current knowledge of  galaxy morphology is based on
the pioneering work of several dedicated observers who
classified thousands of galaxies  and catalogued them.
However, projects  such as the APM/2dF and the Sloan digital
sky surveys will  yield millions of galaxy images and spectra.
Classifying very large data sets is obviously  beyond the
capability  of a single person and
classification problems in Astronomy  call for new approaches
(e.g. Odewhan et al. 1991;  Francis et al. 1992; 
Spiekermann 1992; Storrie-Lombardi et al. 1992;
Doi et al. 1992; Serra-Ricart et al. 1993; Abraham et al. 1994, 1996).
 
Artificial Neural Networks (ANNs)  have recently been utilised in Astronomy
for a wide range of problems, e.g. from adaptive optics to 
galaxy classification 
(for review see Miller 1993 and Storrie-Lombardi \& Lahav 1994).
The ANNs approach should be viewed as a general 
statistical framework, rather than as an esoteric approach.
Some special cases of ANNs are statistics we are all 
familiar with. 
However, the ANNs can do better, by allowing non-linearity. 
Here we  illustrate these points 
by examples from the  problem of morphological  classification
of galaxies, using the ESO-LV (Lauberts \& Valentijn 1989) sample
with 13 parameters and $\sim 5200$ galaxies, as analysed by ANNs
(Storrie-Lombardi et al. 1992; Lahav et al. 1996), 
and for a  sample of $\sim 830 $  APM galaxies (Naim et al. 1995a, 1995b).
We also describe a pilot study of  galaxy spectral classification 
(Folkes, Lahav \& Maddox 1996).
The outline of this article  is as follows.
In \S 2 we present a comparative study between experts, 
in \S 3 we discuss ANNs and their application to the morphological 
classification problem, and in \S 4 we consider spectral classification
of galaxies.

\section{Human Classification of   APM Galaxies}

 The motivation for performing  a
comparison between different experts is two-fold:
(i) to study systematically the degree of agreement and reproducibility
between observers, and 
(ii) to use the human classification as `training sets' for
the Artificial Neural Networks and other automated  classifiers.

We have defined a sample
from the APM Equatorial Catalogue of galaxies
(Raychaudhury et al. 1997)
 selected from IIIaJ (broad blue band) plates taken
with the UK Schmidt telescope at Siding Spring, Australia.
 We chose a subsample of 831  galaxies with
major diameter $D\ge 1.2$ arcmin.
The galaxies were scanned in
raster mode at a resolution of 1 arcsec by
the APM facility at Cambridge.

R. Buta, H. Corwin,  G. de Vaucouleurs, A. Dressler,
J. Huchra and  S. van den Bergh,
kindly classified the {\it same} images on the $T$ system.
Statistically, all  6 experts agreed on the exact $T$-type
for only
8 galaxies out of the 831
(i.e. less than 1 \%).
Agreement between pairs of observers in excess of 80 \%
were obtained only to within 2 types.
For each pair of observers $a$ and $b$ the variance was calculated
(cf. Buta et al. 1994): 
$$
\sigma_{ab}^2 = {1 \over N_{ab} } \sum_i [ T_{a,i} - T_{b,i} ]^2,
\eqno (1) 
$$
where the sum is over 
the $N_{ab}$  galaxies for which both observers gave a classification.
The rms dispersion between  between two  observers who looked
at the {\it same} APM images
is between 1.3 to 2.3  $T$-units,  1.8 on  average. 
\footnote 
{It remains to be tested to what extent an expert reproduces 
his/her own classification.}
Detailed analysis and interpretation of this comparison appear elsewhere
(Lahav et al. 1995, Naim et al. 1995a).
As we show below,
it is  encouraging that the  dispersion
we found between the ANN and an expert
is similar to the dispersion between two human experts.

\section { Automated  Classification  by
Artificial Neural Networks}

The  challenge is to design a computer algorithm which
will reproduce classification
to the same degree a student or a colleague of the human expert can do it.
Such an automated procedure usually involves two steps:
(i) feature extraction from the digitised image, e.g.
the galaxy profile, the extent of spiral arms, the colour of the galaxy,
or an efficient compression of the image pixels
into a smaller number of coefficients (e.g. Fourier
or Principal Component Analysis).
(ii) A classification procedure,  in which a computer `learns' from
a `training set' for which a human expert provided his or her classification.

Artificial Neural Networks (ANNs), originally suggested as simplified models
of the human brain, are  computer algorithms which provide
a convenient general-purpose framework
for classification (Hertz et al. 1991).
ANNs are related
to other statistical methods common in Astronomy and other fields.
In particular  ANNs  generalise 
Bayesian methods, multi-parameter fitting,
Principal Component Analysis (PCA), Wiener filtering
and regularisation methods (e.g. Lahav 1994, Lahav et al . 1996).

\vspace{1cm}
\noindent  {\it 3.1 ANNs as non-linear minimization algorithms }
\vspace{0.5cm}

\noindent 
It is common in Astronomy 
to fit a model with several (or many) free parameters to 
the observations. This regression 
is usually done by means of $\chi^2$ minimization.
A simple example of a `model' is a polynomial with the coefficients as
the free parameters.
Consider now  the  specific problem of morphological classification 
of galaxies. If the type is $T$ 
(e.g. on de Vaucouleurs' numerical system [-6,11])
and we have a set of  parameters ${\bf x}$ (e.g. diameters and colours)
then we would like to find free parameters ${\bf w}$ (`weights') 
such that 
$$
\sigma^2  =  {1 \over N_{gal}} \; \sum_i [T_i - f({\bf w}, {\bf x_i})]^2,
\eqno(2) 
$$
where the sum is over the galaxies, is minimized.
The non-linear function $f({\bf w}, {\bf x})$ represents 
 the `network', which consists of a set of input nodes, 
a set of output nodes and one or more layers of `hidden' nodes
between the input and output layers. 
The `hidden layers' allow curved boundaries around clouds of data
points in the parameter space.
Note the similarity between eq. (2) and eq. (1).
Rather than looking at the variance between two experts, 
we minimize here the variance between the expert and the network.
Commonly $f$ is written as a function of: 
$$
z  = \sum_k  w_k x_k,
\eqno (3) 
$$
where the sum here is over the inputs to each node.
A `linear network' has $f(z)=z$, while a non-linear transfer function
could be a sigmoid $f(z)= 1/[1+ \exp(-z)]$ or $f(z) = \tanh (z)$.
While in most  computational problems we only have 10-1000 nodes,
in the brain there are  $\sim 10^{10}$ neurons, each with
$\sim 10^{4}$ connections.

For a given Network architecture the first step is
the `training' of the ANN.
In this step the weights 
are determined by  minimizing `least-squares' (e.g. eq. 2).
Efficient minimization algorithms include
Backpropagation (Rumelhart, Hinton \& Williams 1986)
and Quasi-Newton (e.g. Hertz et al. 1991).

The interpretation of the output depends on the network configuration.
For example, a single output node  provides an `analog' output (e.g. 
predicting the type or luminosity of a galaxy), while several output nodes
can be used to assign Bayesian probabilities to different classes
(e.g. 5 morphological types of galaxies).


\vspace{1cm}
\noindent  {\it 3.2 The  Bayesian  connection}
\vspace{0.5cm}

\noindent 
A classifier can be formulated
from first principles according to Bayes theorem:
$$
P(T_j|{\bf x}) =  { { P({\bf x} | T_j) \; P(T_j) }
\over { \sum_k P({\bf x} | T_k) \; P(T_k) } }
\eqno (4)
$$
i.e. the {\it a posteriori} 
probability for a class $T_j$ given the parameters vector
${\bf x}$
is proportional to the  probability for data 
given a class (as can be derived 
from a training set) times the {\it prior} probability for a class
(as can be evaluated from the frequency of classes in the training set).
However, applying eq. (4)  requires parameterization of the probabilities 
involved. It is common, although not always adequate, to use
multivariate Gaussians.

It can be shown that the ANN behaves like 
a Bayesian classifier, i.e. the output nodes 
produce  Bayesian {\it a posteriori} probabilities 
(e.g. Gish 1990), although it does not implement 
Bayes theorem directly.
It is reassuring (and should be used as a diagnostic) that 
the sum of the probabilities in an `ideal' network 
add up approximately to unity.
For more rigorous and general  Bayesian approaches for modelling  ANNs see 
MacKay (1992).

\vspace{1cm}
\noindent {\it 3.3 PCA, data compression and unsupervised algorithms}
\vspace{0.5cm}

\noindent 
Principal Component Analysis (PCA) allows
reducing the dimensionality
of the input parameter space. 
A pattern can be thought of as being characterized by a point in an
$M$-dimensional parameter space.  
One may wish a more compact data description, 
where each pattern is described by $M'$ quantities, 
with $ M'\ll M$. This can be 
accomplished by Principal Component Analysis (PCA), a well known statistical 
tool commonly used in Astronomy 
(e.g. Murtagh \& Heck 1987 and references therein).
The PCA method is also known in the literature as
Karhunen-Lo\'eve or Hotelling  transform, 
and is closely related to the technique of Singular Value Decomposition.
By identifying the {\it linear}
combination of input parameters with maximum variance, PCA 
finds $M'$ variables (Principal Components)
that can be most effectively used 
to characterize the inputs. 
PCA is in fact an example of `unsupervised learning', in which an
algorithm or a linear `network' discovers for itself features and
patterns (see e.g. Hertz et al. 1991 for review).  
ANNs can be used to to generate
`non-linear PCA'.
Serra-Ricart et al. (1993) have compared standard PCA to `non-linear
PCA', illustrating how the latter successfully identifies classes
in the data.
Another unsupervised method is Kohonen's Self Organized Map, 
recently used e.g. 
for star/galaxy separation (Mahonen \& Hakala 1995) and galaxy morphological 
classification (Naim et al. 1996). 

\vspace{1cm}
\noindent  {\it 3.4 Results for
galaxy morphological classification by ANNs  }
\vspace{0.5cm}

\noindent 
Storrie-Lombardi et al. (1992)  and  Lahav et al. (1996)
have analysed with ANNs
the ESO-LV (Lauberts \& Valentijn 1989) sample
of about 5200 galaxies, using 13 machine parameters (all scaled to be distance 
independent).
Using a network configuration 13:3:1  (with 46 weights, including `bias') 
for the ESO-LV galaxy data, 
with both the input data and the output $T$-type scaled 
to the range [0, 1] and with sigmoid transfer functions, 
we found dispersion $\Delta T_{\rm rms}  \sim 2$
between the ANN and the experts (LV)
over the $T$-scale [-5, 11].

For a net configuration 13:13:5, where the output layer corresponds 
to probabilities for  
5 broad classes (E, S0, Sa+Sb, Sc+Sd, Irr), 
we found a success rate for perfect match of 64 \%.
Our experiments  indicate that non-linear ANNs 
can achieve better classification than the naive Bayesian classifier 
with Gaussian probability functions, for which the success rate is only 
56 \%. 

Naim et al. (1995b) have applied the same techniques
to the APM sample of 830 galaxies described above, by extracting
features directly from the images, and training the net on the human
classification from the 6 experts.  When the network was trained and
tested on individual expert, the rms dispersion varies between 1.9 to
2.3 $T$-units over the 6 experts.  A better agreement, 1.8 $T$-units,
was achieved when the ANN was trained and tested on the mean type as
deduced from all available expert classifications. 
There is a remarkable similarity in the dispersion between
two human experts and that between ANN and experts !  In other words,
our results indicate that the ANNs can replicate the expert's
classification of the APM sample as well as other colleagues or
students of the expert can do.

\section { Spectral Classification of Galaxies}

Galaxy spectra provide another probe of the intrinsic galaxy
properties.  The integrated spectrum of a galaxy is an important
measure of its stellar composition as well as its dynamical
properties.  Spectra can be obtained to
larger redshifts than morphologies and, as 1-D datasets,
are easier to analyse. 
Apart from its relevance for
environmental and evolutionary studies, new classes of objects may be
discovered as outliers in spectral parameter space.

Although the concept of spectral classification
of galaxies dates from Humason (1936) and Morgan \& Mayall (1957), few
uniform data sets are available and most contain only a small number
of galaxies (e.g. Kennicutt 1992).
  Recent spectral analyses for classification were out carried by
Francis et al. (1992) for QSO spectra, 
von-Hippel et al. (1994) and Storrie-Lombardi et al. (1994) 
for stellar spectra, and in 
particular for galaxy spectra  by 
Connolly et al.(1995) and
Sodr\'e \& Cuevas (1996).

In  a  recent pilot-study Folkes, Lahav \& Maddox (1996) analysed the spectra
of Kennicutt (1992) and simulated the spectra expected to be observed 
with the 2-degree-field (2dF) 400-fibre
facility at the Anglo-Australian Telescope.
PCA was used to compress the spectra from $\sim 800$ wavelength bands 
to $\sim 8$ principal components. The first principal component shows correspondance
to well-known lines (e.g. $H_{\alpha}$, $H_{\beta}$, and $O$II).
The first few principal components provide a useful new compact space
to describe a spectroscopic Hubble sequence.
Moreover,  reconstruction  by using only 8 components 
allows us to  recover successfully the underlying signal 
from  noisy spectra.
It is
also possible to use a sample for which both the $T$-type and the
spectra are available and to train an ANN to predict $T$-type.
In a sense, the ANN provides here mapping between galaxy spectrum and image.
Reliable classification, with more than 90 \% of the normal galaxies 
correctly classified into 5 broad bins, can be expected to the magnitude limit 
of the 2dF survey ($b_J=19.7$).
The ANN classification is more successful than either a $\chi^2$ template
matching approach or a classification based solely on the projection on 
the first principal component.
We intend
to carry out 
automated classification using the above and other methods for the 
 250,000  2dF galaxy spectra.

\section { Discussion}

It is encouraging that in the problem of morphological classification 
of galaxies,  one of the last remaining subjective areas
in Astronomy, ANNs can  replicate 
the classification by a human expert
almost to the same degree of agreement 
as that between two human experts, to within 2 $T$-units.
A pilot study shows that spectral classification 
can be done, with future applications to the 2dF redshift survey.
The challenge for the future is  to develop efficient methods 
for feature extraction and  `unsupervised' algorithms,
combining multi-wavelength information 
to define a `new Hubble sequence' without any prior human
classification.

\section*{Acknowledgments}

I grateful to  S. Folkes, J. Hjorth, S. Maddox, A. Naim, L. Sodr\'e and M.
Storrie-Lombardi for their contribution to the work presented here, 
and for helpful  discussions.

\end{document}